\documentclass[12pt]{iopart}

\usepackage{dcolumn}
\usepackage{graphicx,times} 
\usepackage{makecell}
\usepackage{braket}

\usepackage{color}
\usepackage{cite}
\usepackage{iopams}
\usepackage{multirow}
\usepackage{makecell}
\usepackage[table,xcdraw]{xcolor}
\usepackage[hidelinks]{hyperref}

\usepackage[utf8]{inputenc}

\begin{document}

\title{Phase separation in a mixture of trapped charged Bose-Einstein condensates}

\author{S. Seyyare Aksu~\footnote{Present address: Department of Physics, Koç University, Sariyer 34450, Istanbul, Turkey}}
\ead{seaksu@ku.edu.tr}
\address{Department of Physics, Mimar Sinan Fine Arts University, Bomonti 34380, Istanbul, Turkey}
\author{A. Levent Suba{\c s}{\i}}
\ead{alsubasi@itu.edu.tr}
\address{Department of Physics, Istanbul Technical University, Maslak 34469, Istanbul, Turkey}
\address{Center for Nonlinear Studies, Los Alamos National Laboratory, Los Alamos, New Mexico 87545, USA}
\author{Nader Ghazanfari} 
\ead{nghazanfari@msgsu.edu.tr}
\address{Department of Physics, Mimar Sinan Fine Arts University, Bomonti 34380, Istanbul, Turkey}


\begin{abstract}
We study the phase separation configurations and their rotational properties for a mixture of two interacting charged Bose-Einstein condensates subject to a magnetic field trapped in disc and Corbino geometries. We calculate the ground state energies of azimuthal and radial phase separation configurations using the Gross-Pitaevskii and the Thomas-Fermi approximations. We show that the results for experimentally relevant system parameters from both approaches are in good agreement. The immiscible mixture in both geometries with equal intracomponent interactions favors the azimuthal phase separation for all intercomponent interactions. Only an imbalance in the intracomponent interactions can result in a transition to the radial phase separation, for which the transition becomes sensitive to the shape of the trap. We present phase diagrams as function of the inter and intracomponent interactions. While the radial phase separation is widely favoured in disc geometry, the azimuthal phase separation is favoured for narrower Corbino geometries. We explore the rotational properties of the spatially separated condensates under the magnetic field, studying their angular momenta and velocity fields. The quantization of circulation breaks down for the azimuthal phase separation. In this case, the bulk region of the condensate continues to display superfluid flow behavior whereas the velocity field shows a rigid body behavior along the phase boundaries.
\end{abstract}

\submitto{\jpb}
\maketitle
\section{Introduction}\label{sec:introduction}
Phase separation is a typical aspect of multispecies ultracold atomic systems with repulsive intercomponent interactions. In the immiscible phase the components occupy non-overlapping separate spatial regions. 
The zero temperature mean field studies have revealed that a homogeneous mixture of two miscible Bose-Einstein Condensates (BECs) turns into an immiscible mixture displaying phase separation when intercomponent interaction exceeds the geometric mean of the intracomponent interactions~\cite{Pethick2008}. This condition (with a slight shift) remains valid for relatively large perturbations from uniformity, where the shift is determined by the geometry of the trapping potential~\cite{Smyrnakis2009, Wen2012}.
\par 
The static and dynamic features of phase separations in two component atomic BECs have been studied both experimentally~\cite{Hall1998, Tojo2010, Papp2008, Nicklas2011, Wang2015, McCarron2011, Wacker2015, Burchianti2018, Schulze2018} and theoretically~\cite{Ho1996, Pu1998, Timmermans1998, Ohberg1998, Shi2000, Ao1998, Trippenbach2000, Svidzinsky2003, Shimodaira2010, White2016, Suthar2015, Suthar2016, Bandyopadhyay2017, Mason2013, Lee2016, Lous2018, Abad2014, Roy2015, Chen2019, Richaud2019, Hejazi2020}. Experiments have been carried out for mixtures with two different atomic species~\cite{McCarron2011, Wacker2015, Wang2015}, different isotopes of the same atoms~\cite{Papp2008} or different hyperfine states of the same isotopes~\cite{Hall1998, Tojo2010}. Theoretical studies have mainly been performed at the mean-field level for trapped atoms using the Thomas-Fermi (TF)~\cite{Ho1996, Pu1998, Timmermans1998}, the Gross-Pitaevskii (GP)~\cite{Ao1998, Trippenbach2000, Svidzinsky2003, Shimodaira2010, White2016, Bandyopadhyay2017}, or the Bogoliubov-de Gennes approaches~\cite{Abad2014, Roy2015, Chen2019}. These studies have focused on the transition from the miscible to the immiscible state and on the physical properties of immiscible states for both non-rotating and rotating BECs. 
\par
The phase separation configurations are determined by the difference in the strength of the intracomponent interactions and the shape of external potential~\cite{Svidzinsky2003, Shimodaira2010, Mason2013, Abad2014}. For a mixture in a toroidal trap two configurations of phase separation can occur: the azimuthal phase separation (APS) and the radial phase separation (RPS) for which the components are restricted to semi-circular and concentric full circular non-overlapping annular regions, respectively (see Fig.~\ref{fig:density-distributions}).
The APS is the ground state of symmetric immiscible mixtures, i.e. equal particle masses and equall intracomponent interaction energies~\cite{Shimodaira2010, Mason2013, Abad2014}.
However, a phase transition from APS to RPS occurs by introducing an imbalance in the system~\cite{Abad2014}.  
\par
In addition to the spatial separation of the density distributions, the phase separation also affects other physical properties of the condensates. The rotational properties of the APS configurations show different behaviour with respect to those of the RPS. While the circulation of the velocity field for both condensates remains quantized in an RPS, it breaks down for an APS. The angular momenta of both condensates exhibit a smooth transition from quantized to continuous values as the mixture is driven through a transition from the RPS to the APS~\cite{White2016}. 
\par
The mixture of BECs have been studied and realized in different geometries \cite{White2016, Abad2016, Mason2013, Ryu2014, Eckel2014, Moulder2012, Beattie2013, Chen2019, Ghazanfari2014, Ramanathan2011, Murray2013, Beattie2013, Burchianti2018, Schulze2018, Ryu2013, Aksu2019} including toroidal trapping potentials where the fluid can be modeled as a one-dimensional system on a ring geometry~\cite{Chen2019, Aksu2019, Abad2016, Eckel2014, Moulder2012, Murray2013} or two dimensional system on a toroidal/Corbino~\cite{White2016, Mason2013, Ryu2014, Beattie2013, Ramanathan2011, Ryu2013} and disc geometries~\cite{Wacker2015, McCarron2011, Burchianti2018, Schulze2018, Ghazanfari2014}. 
In current experiments for ultracold atomic and molecular systems, different trap geometries can be generated and the interaction strengths can be finely tuned~\cite{Bloch2008}. 
Moreover, artificial magnetic fields for ultracold gases can make neutral atoms behave as if they are electrically charged~\cite{Lin2009a, Lin2009b}. The strength of this magnetic field depends on the internal structure of atoms and thus can be species selective. This allows us to consider various mixtures of synthetically charged superfluids~\cite{Unal2016, Aksu2019, Subasi2020, Garaud2014, Hejazi2020}. Therefore, two condensates with equal or different rotation frequencies can be created.
\par
In this study we consider phase separated mixtures of two interacting charged BECs subject to a weak magnetic field and trapped in disc and Corbino geometries. We analyse the conditions for the phase transition between the mentioned configurations of the immiscible phase and study their rotational properties. We explore the phase separated mixtures with different inter and intracomponent interactions and comment on the effects of charge imbalance at this level of approximations. We discuss how the shape of the trap becomes relevant when there exists an asymmetry between the physical properties of the components.
\par
We use both the GP and the TF approximations to investigate the ground state and the rotational properties of mixtures as a function of both inter and intracomponent interactions and the applied magnetic field. The coupled GP equations describing the system are solved using the imaginary time evolution~\cite{Castin1999}. We compare the results of the GP simulations with solutions obtained from the TF approximation improved by a variational interface energy and conclude that the latter works reasonably well for experimentally relevant systems.
\par
This article is organized as follows: In the next section we define the physical properties of a mixture of two synthetically charged Bose-Einstein condensates in a harmonic trap and subject to an artificial magnetic field. We provide the equations describing the BEC mixture within the GP and the TF approximations. In Sec.~\ref{sec:balanced-case}, we consider the case with equal intracomponent interactions. The ground state of an immiscible mixture has the APS configuration for any value of the inter-component interaction and synthetic charges. We also show that for weak magnetic fields and synthetic charges considered, the resulting kinetic energy does not play a significant role in determining the phase boundary. In Sec.~\ref{sec:imblalanced case}, we present phase diagrams showing the phase separation configurations for different intra and intercomponent interactions where the interface energy plays a decisive role. In Sec.~\ref{sec:rotational-properites} we analyze the rotational properties of the condensates and finally in Sec.~\ref{sec:Conclusion} we summarize and discuss our results.
\section{Mixture of two charged superfluids}\label{sec:Mixture of charged Bose-Einstein condensates}
We consider a mixture of two charged superfluids consisting of equal number of atoms $N_1=N_2=N$, with the same particle masses $M$, and synthetic charges $q_1$ and $q_2$. The mixture is strongly confined along the longitudinal direction, $z$, in a harmonic potential of the form 
\begin{equation}
    V_\mathrm{ext}(\mathbf{r})=V_\perp+V_z=\frac{1}{2}M\omega_\perp^2 \left(\textbf{r}_\perp-\textbf{r}_{\perp0}\right)^2 + \frac{1}{2}M\omega_z^2 z^2
\end{equation}
where $\omega_z \gg \omega_\perp$ ($\omega_\perp=\omega_x=\omega_y$) are the trapping frequencies , and $r_\perp=\sqrt{x^2+y^2}$ denotes the radial distance in the $xy$-plane. We adapt an effective two-dimensional description. For $\textbf{r}_{\perp0}=0$ this potential gives a disc geometry and for a finite $\textbf{r}_{\perp0}$ it gives a Corbino geometry~\cite{Bargi2010}. 
The system is under a uniform artificial magnetic field ${\bf B}=B\hat{e}_z$ along the $z$-axis generated by the symmetric vector potential $\mathbf{A}(\mathbf{r})=\frac{B}{2}(-y,x,0)$ in the Coulomb gauge. 
\par
For each species $j=1,2$ the single-particle Hamiltonian in a two-dimensional harmonic potential can be written as
\begin{eqnarray}
    H_j &=& \frac{1}{2M}\left[ \mathbf{p}-q_j \mathbf{A}(\mathbf{r}) \right]^2 +  V_\perp(\mathbf{r}),\\\nonumber
      &=& \frac{p^2}{2M}-\omega_j (xp_y-yp_x) + \frac{1}{2}M\omega_j^2 r_\perp^2 + V_\perp(\textbf{r}),
\end{eqnarray}
where $\omega_j=q_j B/2M$ are the cyclotron frequencies. Both species have the same particle mass but may feel different trapping sizes due to the additional term coming from minimal coupling, i.e. $\frac{1}{2}M\omega_j^2 r_\perp^2$ if they have different synthetic charges. 
\par
The two-dimensional intracomponent and intercomponent interactions, are modelled by the short-range contact interaction with coupling constants defined by
\begin{eqnarray}
    g_{jk} = \frac{\sqrt{8\pi}\hbar^2a_{jk}}{Ml_z}.
\end{eqnarray} 
where $j,k = \{1, 2\}$ enumerate components in the mixture, $a_{jk}$ denote the three dimensional s-wave scattering lengths and $l_z=\sqrt{\hbar/M\omega_z}$. 
\par
The time evolution of the trapped interacting condensates is governed by a pair of coupled GP equations 
\begin{eqnarray}
    \label{disc-coupld-GPE}\nonumber
     i\partial_t \psi_j(\textbf{r},t)\!\! &=&\!\! \left[-\frac{\nabla^2 }{2}- \Omega_j L_z\! + \! \frac{1}{2}\Omega_j^2 r^2\!+\! \frac{1}{2} (\textbf{r}-\textbf{r}_0)^2\right]\!\psi_j(\textbf{r},t)\\ 
    &+& \sum_{k=1,2}U_{jk}\vert\psi_k(\textbf{r},t)\vert^2\psi_j(\textbf{r},t).
\end{eqnarray}

\begin{figure}[t]
\centering
\resizebox{.6\columnwidth}{!}{%
\includegraphics{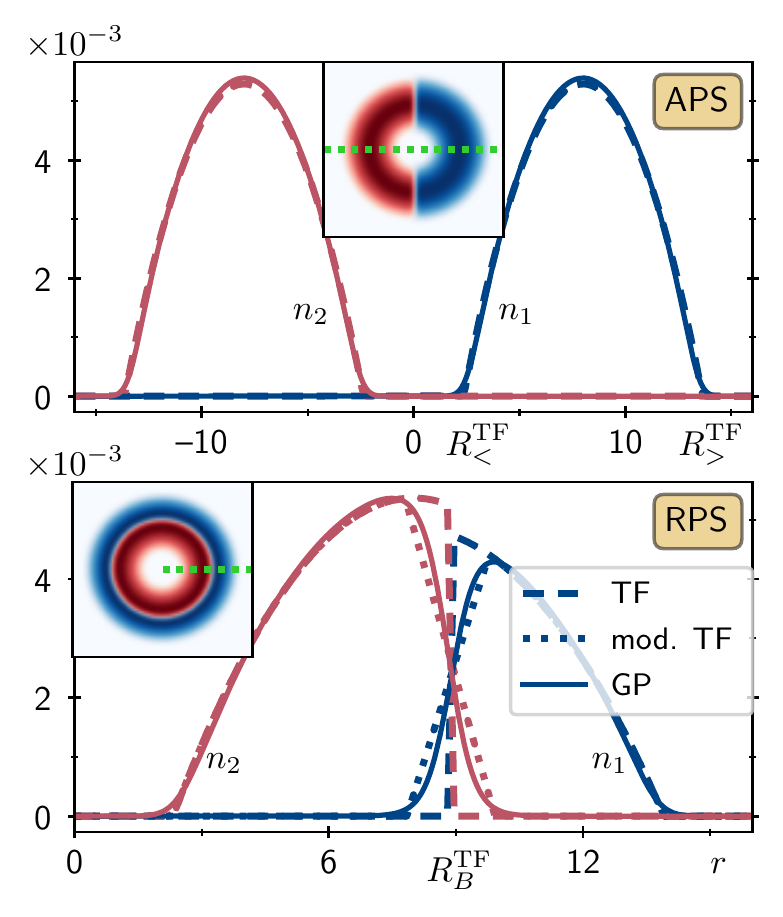}
}
\caption{Density distributions of two phase separated uncharged BECs in the APS (top panel) and in the RPS (bottom panel) configurations. 
The results of the GP calculations (solid lines) are in good agreement with those of both bare and modified TF approximations (dashed lines). The modified TF approach includes a variational interface energy obtained with a linear density profile across the boundary.
The insets show the schematic two-dimensional component densities and the dotted (green) line is the cross-section along which the profiles are shown. The parameter values are $U_{12}=1.2U_1=1.2U_2$ for APS and $U_{12}=U_1=1.2U_2$ for RPS with $r_0=8$ and $U_2=3000$ being common to both.
Quantities in all figures are plotted in dimensionless units.
\label{fig:density-distributions}
}
\end{figure}
All the relevant quantities are made dimensionless via scaling lengths by the oscillator length $l_\perp=\sqrt{\hbar/M\omega_\perp}$, time by $1/\omega_\perp$, angular momenta by $\hbar$, and order parameters $\psi_j$ by $\sqrt{N}/l_\perp$ so that $r=r_\perp/l_\perp$, $\Omega_j=\omega_j/\omega_\perp$, $U_{jj} \equiv U_j=\sqrt{8\pi}Na_{jj}/l_z$ and $U_{12}=\sqrt{8\pi}Na_{12}/l_z$. 
Here, the condensate wave functions are normalized via $\int d{\bf r} |\psi_j ({\bf r},t)|^2 = 1$ and the dimensionless angular momentum $ L_z=\left(xp_y - yp_x\right)/\hbar$, which is given as $L_z=-i \partial / \partial \theta$ in polar coordinates. Note that the information about the charge of each condensate is embedded inside the rotation frequency $\Omega_j$. Thus the effect of charge imbalance ($q_1\neq q_2$) can be interpreted as applying componentwise rotations to the condensates, i.e. $\Omega_1\neq \Omega_2$.
\par
Beside the GP approach the TF approximation is also used to study the properties of BECs when the kinetic energy can be neglected compared with the interaction energies.
The TF approximation provides algebraic equations to examine the system and is valid when $Na_{ij}/l_\perp \gg 1$. In this study, at least one of the components is subject to a magnetic field. 
In order to account for the superfluid flow characteristics within the TF approximation we make the ansatz for the condensate wavefunctions
$\psi_j(\textbf{r},t)=\phi(r)e^{il_j\theta}e^{-i\mu t}$ and only neglect the derivatives with respect to the radial coordinate. 
Here $\mu$ is the chemical potential and  $l_j$ is the variational parameter of the TF ansatz which gives the angular momentum of component $j$. 
While for the RPS, $l_j$ can only be an integer, for the APS, $l_j$ can take non-integer values.
Keeping the derivative with respect to the azimuthal angle, we write the TF equations as
\begin{equation}
\left(
\begin{array}{c}
n_1 \\
n_2 
\end{array} 
\right)
= 
\frac{1}{U_1 U_2-U_{12}^2}
\left(
\begin{array}{cc}
U_{2} & -U_{12} \\
-U_{12} & U_{1} 
\end{array}  
\right)
\left(
\begin{array}{c}
\varepsilon_1  \\
\varepsilon_2
\end{array}  
\right),
\end{equation}
where $\varepsilon_j=\mu_j - \frac{1}{2}\Omega_j^2 r^2 - \frac{1}{2}\left(r-r_0\right)^2  + \Omega_j l_j - \frac{l_j^2}{2r^2}$. Note that the chemical potentials $\mu_j$ adjust the total particle numbers and $n_1,n_2\ge 0$ determine the TF radii $R^\mathrm{TF}$.
\par 
When phase separation takes place ($U_{12}^2>U_1U_2$) the components mostly occupy non-overlapping spatial regions. We further 
simplify the TF equations in these cases by assuming either an azimuthal or a radial boundary between the components. In this way, by assuming strictly non-overlapping components the above TF equations decouple leading to the corresponding one-component TF equations $n_j=\varepsilon_j/U_j$ subject to the given boundary. Next, we optimize the boundary assuming zero interface energy at the boundary.
The resulting APS and RPS configurations can have very close energies and in order to accurately decide on the ground state configuration, the interface energy at the boundary between the components must be taken into account. Having obtained the TF solutions for each component we calculate the interface energy within the local density approximation in the following way. 
\par 
The interface energy involves contributions from both interactions and confinement at the interface. 
Within the TF approximation in the immiscible phase the boundary region can be defined as the region where the condensate wavefunctions overlap and recover from zero to their bulk densities. The relevant length scales for the width of the boundary are the dimensionless healing lengths of the condensates $\xi_j=\sqrt{1/4U_jn_j}$ as well as the penetration depths $\Lambda_j = \xi_j / \sqrt{U_{12}/\sqrt{U_1U_2}-1}$~\cite{Ao1998}.
\par
Depending on the values of the intra and intercomponent interactions, the shape of the condensate wavefunctions can take different forms at the boundary.~\cite{Schaeybroeck2008}. 
The shape of the boundary region has been studied in multiple limiting values of the above healing and penetration lengths covering most of the parameter space~\cite{Ao1998, Timmermans1998, Barankov2002, Schaeybroeck2008, Indekeu2015}.
The governing dimensionless parameters can be identified as $K=\left(\xi_j/\Lambda_j\right)^2=U_{12}/\sqrt{U_1U_2}-1$ (for equal masses) 
and $\xi_2/\xi_1=(U_1/U_2)^{1/4}$ (at balanced bulk pressures).
The parameters used in our calculations which are motivated by the experimental values fall in between the carefully studied regions. We, therefore, adopt an earlier physically motivated variational approach for the interface energy~\cite{Timmermans1998}. 
\par
We assume that the width of the boundary region is locally determined by the optimization of the boundary energy density. Taking the kinetic and interaction energies into account, the pressure balance can be satisfied by making an ansatz for the boundary of width $b$ over which the densities vary linearly. (See Fig.~\ref{fig:density-distributions} for the modified TF density profile.) The kinetic energy density is approximated by $n_j/(2b^2)$ and $b$ is obtained by minimizing the energy density locally across the boundary as $b^* = 2\xi_1 \sqrt{3\left(1+\sqrt{U_1/U_2}\right)/K} $. In this way, the interface energy density can be written as~\cite{Timmermans1998}
\begin{equation}
    \sigma(r) = \xi_1(r) P_1(r) \Sigma
\end{equation}
where the healing length $\xi_1$ and the pressure $P_1(r) = U_1 n_1^2(r)/2$ are evaluated locally along the boundary and 
$\Sigma = 4\sqrt{K\left(1+\sqrt{U_1/U_2}\right)/3}$. 
\par 
Finally, the total interface energy is obtained by integrating the above energy density over the boundary region. For the angular and radial separation, we obtain the following expressions:
\begin{eqnarray}
    E_s^\mathrm{APS} &=& 2 \int_{R_<^\mathrm{TF}}^{R_>^\mathrm{TF}}  \sigma(r) \mathrm{d}r  \\
    E_s^\mathrm{RPS} &=& 2\pi R_B^\mathrm{TF} \sigma(R_B^\mathrm{TF})
\end{eqnarray}
where $R_<^\mathrm{TF},R_>^\mathrm{TF}$ denote the limits of the radial extend of the boundary in the APS configuration and $R_B^\mathrm{TF}$ is the radius of the circular boundary in the RPS configuration (see Fig.~\ref{fig:density-distributions}). We find that complementing the TF energy with the above interface energy gives accurate results when compared with our numerical simulations of the GP equations. 

\begin{table}[t]
\centering
\begin{tabular}{ l | c  c  c  c  c  c  }
        & $(\Omega_1,\Omega_2)$     & $E_{kin}$   & $E_{pot}$ & $E_{int}$ & $E^{12}_{int}$ & $E_{tot}$  \\ 
\hline
\hline
APS(GP) & (0.01,0.01)                 &  0.091     & 6.668    & 13.664    & 0.155          & 20.578        \\ 
RPS(GP) & (0.01,0.01)                 &  0.224     & 6.814    & 12.894    & 0.914          & 20.845      \\ 
APS(TF) & (0.01,0.01)                 &  0.037     & 6.815    & 13.633    & 0.035          & 20.521       \\ 
RPS(TF) & (0.01,0.01)                 &  0.201     & 6.809    & 13.639    & 0.200          & 20.848      \\
\hline   
APS(GP) & (0.01,0)                   & 0.091    & 6.668      & 13.664     & 0.155          & 20.578    \\ 
RPS(GP) & (0.01,0)                   & 0.224    & 6.814      & 12.894     & 0.914          & 20.845     \\ 
APS(TF) & (0.01,0)                   & 0.037    & 6.815      & 13.632     & 0.035          & 20.519  \\ 
RPS(TF) & (0.01,0)                   & 0.200    & 6.809      & 13.638     & 0.200          & 20.848    \\
\end{tabular} 
\caption{
Contributions to the total energy for mixtures of equally and unequally charged BECs trapped in a Corbino geometry with an inner radius of $r_0=12$ and an equal intracomponent interaction $U_{1}=U_{2}=U=5000$ and intercomponent interactions $U_{12}=1.2U=6000$. The energy values calculated by GP and TF approximations are in a good agreement. Azimuthal phase separation is favoured for both equally and unequally charged mixtures.}
\label{table:sym-corbino}
\end{table}

\par
We use the XMDS2 software package for imaginary time evolution simulations~\cite{xmds2}. The XMDS2 library can solve systems of initial-value partial and ordinary differential equations. It can simulate a system of equations in arbitrary number of dimensions.
For time evolution one can choose different algorithms such as Runge-Kutta and adaptive Runge-Kutta algorithms. 
We employ a split-step time evolution where the kinetic energy operators involving the derivatives are evaluated in the Fourier space.
We run fourth-fifth order adaptive Runge-Kutta algorithm in order to find the stationary ground state solution of a mixture of charged BECs in two dimensions by employing imaginary time evolution with the following typical parameters. 
For example, we choose a square grid of size $512\times 512$ with the spatial extent $[-20,20]$ for our simulations in Fig.~5. 
The grid spacing is approximately $\Delta x = 2 \times 20 / 512 = 0.078$. The healing length of the system is $\xi = 1/\sqrt{gn}$, thus taking density values around the phase boundary, the magnitude of the healing length of the system is $\xi \approx 0.289$.  
Similarly for Fig.~6, we choose a square grid of size $512\times 512$ with the spatial extent $[-12,12]$ such that the grid spacing is approximately $\Delta x = 2\times 12 / 512 = 0.047$ and the magnitude of the healing length of the system is $\xi \approx 0.154$. 
The grid spacing and the healing length should at least be comparable in order to capture the changes in the wave function accurately. 
For example, the ratio of the healing length to grid spacing for the interaction $U=3000$ is $\xi/\Delta x = 3.695$ and $\xi/\Delta x = 3.292$, for Corbino geometry in Fig.~5 and disc geometry in Fig.~6, respectively. 

The convergence is assessed by monitoring the change of the wavefunction during the imaginary time evolution. We calculate the difference $1- |\braket{\psi_\tau |\psi_{\tau+\Delta\tau}} |/| \braket{\psi_\tau|\psi_\tau} |$ using wavefunction at times $\tau$ and $\tau+\Delta\tau$ and typically require it to be less than $10^{-6}$ for imaginary time interval $\Delta\tau \sim 5-10$ in units of $1/\omega_\perp$.

\par
We study the phase separation configurations in two cases, namely the interaction-balanced case for which intracomponent interactions are taken equal, i.e. $U_1=U_2=U$ and the interaction-imbalanced case for which $U_1 \neq U_2$. We provide comparison of various energy expectation values from these approaches and obtain the phase separation configurations based on both GP and TF results in the following sections. For the sake of simplicity, in the following sections we use the TF abbreviation for the modified TF. 

\begin{table}[t]
\centering
\begin{tabular}{ l | c  c  c  c  c  c  }
        & $(\Omega_1,\Omega_2)$     & $E_{kin}$   & $E_{pot}$ & $E_{int}$ & $E^{12}_{int}$ & $E_{tot}$ \\ 
\hline
\hline
APS(GP) & (0.01,0.01)                 &  0.216     & 36.967    & 37.570   & 0.869         & 75.622         \\ 
RPS(GP) & (0.01,0.01)                 &  0.330     & 37.034    & 36.983   & 1.504         & 75.651               \\ 
APS(TF) & (0.01,0.01)                 &  0.196     & 37.611    & 37.614   & 0.193         & 75.614        \\ 
RPS(TF) & (0.01,0.01)                 &  0.333     & 37.702    & 37.523   & 0.330         & 75.889             \\
\hline   
APS(GP) & (0.01,0)                   & 0.216    & 36.967      & 37.570   & 0.869         & 75.621  \\ 
RPS(GP) & (0.01,0)                   & 0.329    & 37.050      & 36.982   & 1.504         & 75.850   \\ 
APS(TF) & (0.01,0)                   & 0.195    & 37.612      & 37.613   & 0.192         & 75.613   \\ 
RPS(TF) & (0.01,0)                   & 0.332    & 37.702      & 37.523   & 0.330         & 75.888   \\
\end{tabular}
\caption{
Contributions to the total energy for mixtures of equally and unequally charged BECs trapped in a disc geometry with equal intracomponent interaction $U_{1}=U_{2}=U=5000$  and intercomponent interactions $U_{12}=1.2U=6000$. The energy values calculated by GP and TF approximations are in a good agreement. Azimuthal phase separation is favoured for both equally and unequally charged mixtures.}
\label{table:sym-disc}
\end{table}
\begin{table}[hb]
\centering
\begin{tabular}{ l | c | c | c | c}
$r_0$      & $\Omega$ &  \thead{$U\!=\!1000$ \\ \begin{tabular}{c c} APS & RPS \end{tabular}} & \thead{$U\!=\!10000$ \\ \begin{tabular}{c c} APS & RPS \end{tabular}}  & \thead{$U\!=\!15000$ \\ \begin{tabular}{c c} APS & RPS \end{tabular}} \\
\hline
\hline
0          & 0.01 & \begin{tabular}{c c} 34.028 & 34.303 \end{tabular} 
           & \begin{tabular}{c c} 106.775 & 107.051 \end{tabular} & \begin{tabular}{c c} 130.686 & 130.959 \end{tabular} \\

8          & 0.01 & \begin{tabular}{c c} 9.235 & 9.566 \end{tabular}  
           & \begin{tabular}{c c} 42.691 & 42.979 \end{tabular} & \begin{tabular}{c c} 55.903 & 56.183 \end{tabular} \\

12         & 0.01 & \begin{tabular}{c c} 7.035 & 7.384 \end{tabular} 
           & \begin{tabular}{c c} 32.552 & 32.868 \end{tabular} & \begin{tabular}{c c} 42.640 & 42.949 \end{tabular}  \\
\hline 
0          & 0.03 & \begin{tabular}{c c} 34.039 & 34.304 \end{tabular} 
           & \begin{tabular}{c c} 106.812 & 107.063 \end{tabular} & \begin{tabular}{c c} 130.730 & 130.969 \end{tabular} \\

8          & 0.03 & \begin{tabular}{c c} 9.245 & 9.568 \end{tabular}
           & \begin{tabular}{c c} 42.741 & 42.995 \end{tabular} & \begin{tabular}{c c} 55.967 & 56.204 \end{tabular} \\

12         & 0.03 & \begin{tabular}{c c} 7.042 & 7.386 \end{tabular} 
           & \begin{tabular}{c c} 32.587 & 32.878 \end{tabular} & \begin{tabular}{c c} 42.686 & 42.963 \end{tabular}  \\
\hline 
0          & 0.06 & \begin{tabular}{c c} 34.077 & 34.320 \end{tabular}
           & \begin{tabular}{c c} 106.932 & 107.103 \end{tabular} & \begin{tabular}{c c} 130.878 & 131.018 \end{tabular} \\

8          & 0.06 & \begin{tabular}{c c} 9.279 & 9.578 \end{tabular}
           & \begin{tabular}{c c} 42.901 & 43.046 \end{tabular} & \begin{tabular}{c c} 56.170 & 56.269 \end{tabular} \\

12         & 0.06 & \begin{tabular}{c c} 7.154 & 7.394 \end{tabular}  
           & \begin{tabular}{c c} 32.771 & 32.918 \end{tabular} & \begin{tabular}{c c} 42.901 & 43.019 \end{tabular}  \\
\end{tabular}
\caption{Total energies for a mixture of equally charged BECs trapped in a two dimensional harmonic trap with $U_{12}=1.2 U$ for different inner radii $r_0$, intracomponent energies $U_{1}=U_{2}=U$ and rotation frequencies $\Omega_1=\Omega_2=\Omega$ calculated by the GP approximation. The energy values calculated by TF approximations also show that azimuthal phase separation is always favoured for a mixture of BECs.}
\label{table:sym-equally-charged}
\end{table}
\section{Interaction-balanced mixture}\label{sec:balanced-case}
We start with a mixture of two BECs subject to a magnetic field with equal intracomponent interactions, i.e. $U_{1}=U_{2}=U$. This limit can be obtained by having equal number of particles in each gas, $N_1=N_2$, and equal $s$-wave scattering lengths, $a_{11}=a_{22}$, or by choosing $N_1/N_2=a_{22}/a_{11}$. In this case we observe that for both Corbino and disc geometries, the APS is energetically favourable compared to the RPS~\cite{Shimodaira2010, Mason2013, Abad2014}.
\par
We calculate the APS and the RPS energies with the intracomponent interaction strength of $U= 5000$ for both equally and unequally charged cases. An intracomponent interaction strength of $U = 5000$ with $N = 3-4\times 10^5$ particles corresponds to an s-wave scattering length of $5\,$nm, which is reasonable for experimental setups~\cite{Hall1998, Mertes2007, Beattie2013}.
The detailed energy values are given in Table~\ref{table:sym-corbino} and Table~\ref{table:sym-disc}, for a Corbino with $r_0=12$ and a disc geometry, respectively ($E_{kin}$, $E_{pot}$, $E_{int}$, $E^{12}_{int}$, and $E_{tot}$ stand for kinetic, potential, intracomponent, intercomponent and total energies, respectively). 
\par 
We observe that the total APS energy is lower than that of the RPS for both geometries and also for equally and unequally charged cases. 
The results of the TF approximation are in good agreement with those of the GP approach, both qualitatively and quantitatively. (We show examplary density profiles in Fig.~\ref{fig:density-distributions}.) The APS is favourable for weak and strong intracomponent interactions as seen from Table~\ref{table:sym-equally-charged}. We present the APS and the RPS energies for $U=$ 1000, 10000 and 15000 and also for larger rotational frequencies. For a slightly higher magnetic field or an arbitrary charge imbalance the difference between the APS and the RPS energies stays almost unchanged, since the contribution of the rotational kinetic energy is very small. 
\par
\begin{figure}[t]
\centering
\resizebox{.5\columnwidth}{!}{%
\includegraphics{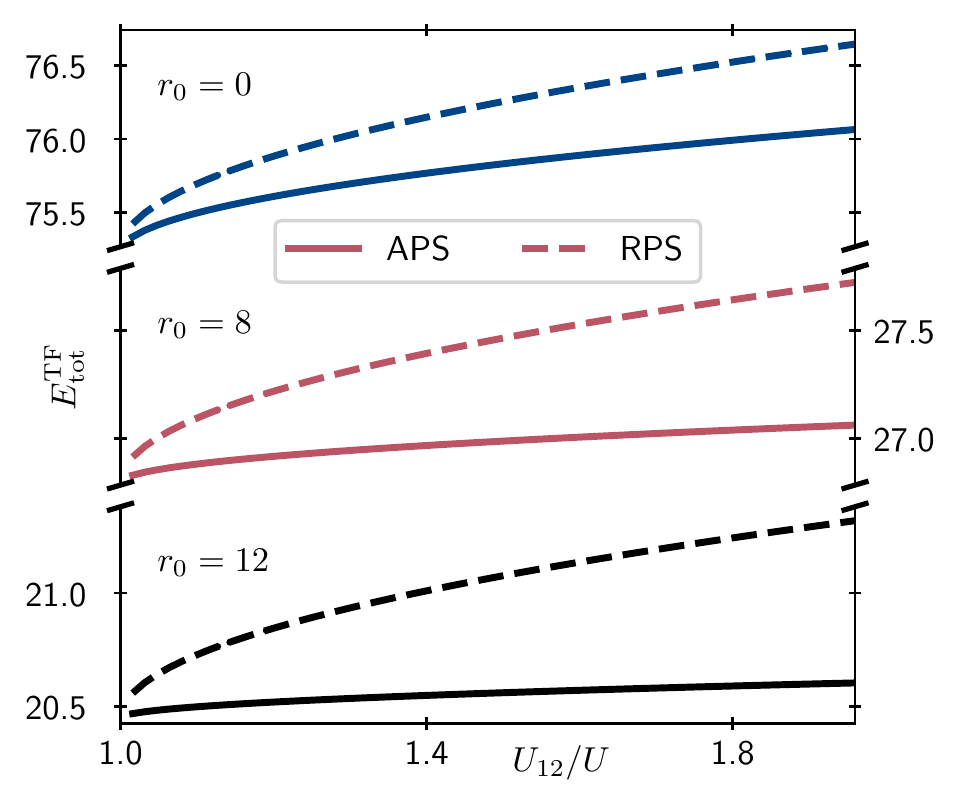}
}
\caption{Total TF energies calculated at fixed intracomponent energies $U_1=U_2=U=5000$ for disc geometry with $r_0=0$, and two Corbinos geometries with inner radii $r_0=8$ and $r_0=12$ as function of the intercomponent energy. The ground state energy of the APS configuration (solid lines) is lower than that of the the RPS configuration (dashed lines) for all values of the interacomponent energy, and becomes energetically more advantageous by increasing $U_{12}$.
\label{fig:energies-symmetric}
}
\end{figure}
A change in the external trapping potential, i.e a change in the shape of the trap geometry, affects the densities and is therefore qualitatively similar to changing the interactions. With equal intracomponent interactions for both components we find the APS configuration energetically favorable for all geometries considered here.
\par
From surveying the magnitude of the energies given in Table~\ref{table:sym-corbino} and Table~\ref{table:sym-disc} it is seen that the difference between the APS and RPS total energies is smaller than the difference between the interface energies in both of our approaches.
When the interface energy $E_s=E_{kin}+E_{int}^{12}$ is neglected, the RPS configuration has lower energy because of its larger boundary compared to that of the APS configuration. 
Even though the interface energy is quantitatively smaller in the TF approximation, it is qualitatively accurate.
Therefore, the configuration of the phase separation is decided by the magnitude of the interface interaction. In other words, throughout a phase separation condensates with the same properties tend to minimize the contribution of boundary effects to the total energy~\cite{Svidzinsky2003}. In the interaction-balanced case, there is no physical factor between BECs to force an unequivalence in density distributions.
\par
The energy difference between the APS and the RPS increases with increasing the intercomponent interactions which is shown in Fig~\ref{fig:energies-symmetric}. We calculated the total energy for a disc geometry in Fig.~\ref{fig:energies-symmetric} with $r_0=0$ and two Corbino geometries with $r_0=8$ and $r_0=12$ for a fixed intracomponent interaction of $U=5000$. In all figures the APS remains the energetically advantageous configuration in an immiscible mixture of BECs with balanced intracomponent interactions. Again, we note that the difference in total energy between the APS and RPS is due to the difference in the interface boundary. A large boundary means a large interface energy. Thus, to obtain the RPS configuration an imbalance between the intracomponent interactions is needed. 
\begin{figure}[t]
\centering
\resizebox{.5\columnwidth}{!}{%
\includegraphics{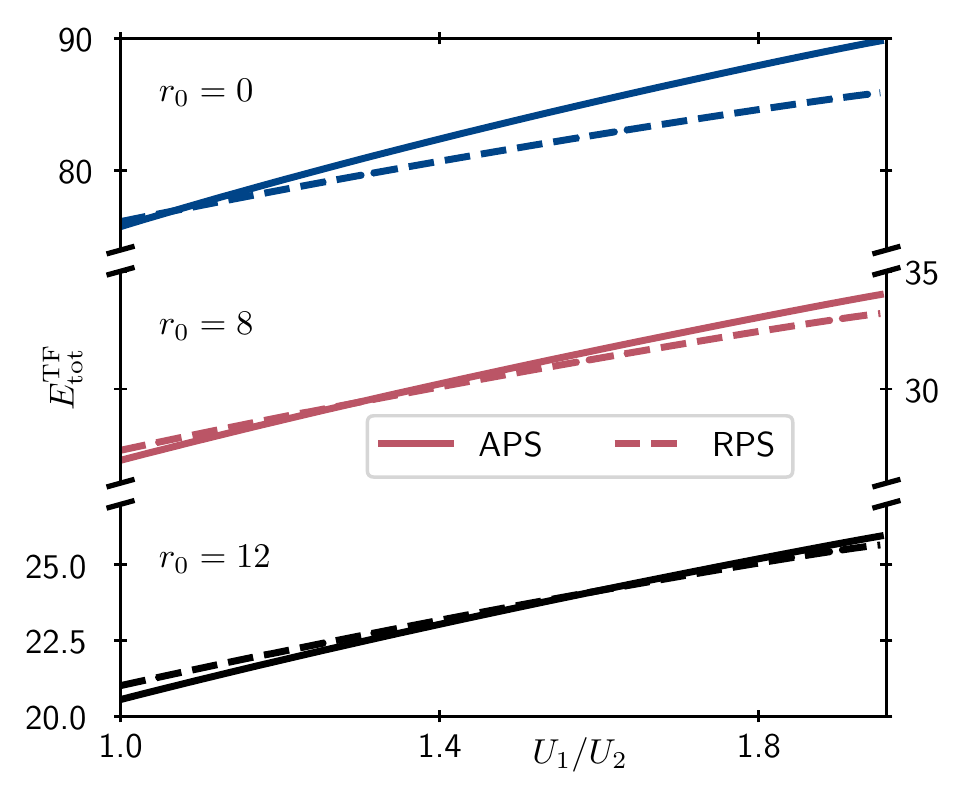}
}
\caption{Total TF energies calculated for APS and RPS configurations for a disc geometry with $r_0=0$ and two Corbino geometries with inner radius $r_0=8$ and $r_0=12$ with interacomponent interaction $U_{12}=7000$ and intracomponent interaction $U_1=5000$. For the disc geometry the transition from APS to RPS happens almost as soon as a small imbalance in the intracomponent interactions is introduced.
\label{fig:ETFvsgamma0}
}
\end{figure}
\section{Interaction-imbalanced mixture}\label{sec:imblalanced case}
\begin{figure*}[ht]
\centering
\resizebox{\textwidth}{!}{%
\includegraphics{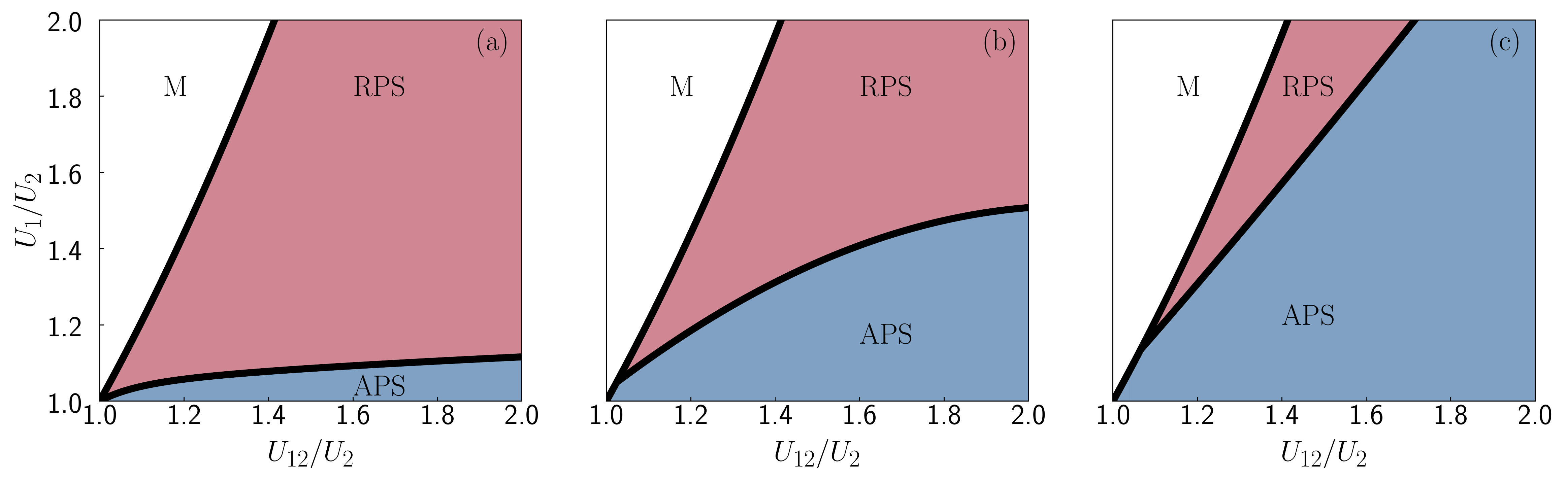}
}
\caption{Phase diagrams as a function of the intercomponent and intracomponent interactions for (a) $r_0=0$ disk geometry, (b) $r_0=8$ and (c) $r_0=12$ Corbino geometries. 
The boundaries for miscible mixture (M) has been calculated analytically according to the condition $U_1U_2 \leq U_{12}^2$. The APS-RPS transition happens in the interaction-imbalanced cases and unlike the interaction-balanced case its boundary is sensitive to the geometry of the trap. The RPS is widely favoured in disc geometry and the APS is favoured for narrower Corbino geometries.
\label{fig:phasediagram}
}
\end{figure*}
In the interaction-imbalanced case for which $U_1\neq U_2$, the phase separation configuration is determined by the difference in the strength of the intracomponent interactions and the shape of the external potential~\cite{Shimodaira2010, Abad2014, Mason2013}.
Fig.~\ref{fig:ETFvsgamma0} exhibits the transition from APS to RPS as a function of the imbalance in the intracomponent interactions. The total energies are calculated by the TF approximation for the intercomponent interaction $U_{12}=7000$ and intracomponent interaction $U_2=5000$. For the disc geometry shown in Fig.~\ref{fig:ETFvsgamma0} ($r_0=0$), the transition from APS to RPS happens as soon as a small imbalance in the intracomponent interactions is introduced. The similar transition for Corbino geometries occurs at larger differences as shown in Fig.~\ref{fig:ETFvsgamma0} with respect to the inner radius of the Corbino, for $r_0=8$ and $r_0=12$, respectively. 
\par 
By a transition from the APS to the RPS, the component with the weaker intracomponent interaction occupies the inner disc or annular region and the strongly interacting one occupies the outer annular region. In this configuration, in order minimize the total energy, the condensate with smaller intracomponent interaction stays in a region with higher density, while the other condensate moves radially outwards reducing its density~\cite{Svidzinsky2003}. The pressure balance in this configuration results in lower total energy despite the fact that the RPS has a larger boundary compared to the APS (see table \ref{table:imbalanced}). For the RPS, the energy loss through the interface energy is compensated by the gain from intracomponent interaction energy minimization, which is a result of components' spatial arrangement.

\begin{table}[ht]
\begin{center}
\begin{tabular}{ l | c  c  c  c  c  c  }
        & $r_0$     & $E_{kin}$    & $E_{pot}$ & $E_{int}$ & $E^{12}_{int}$ & $E_{tot}$  \\ \hline\hline
APS(GP) & 0         &  0.174     & 38.611    & 38.891     & 1.282          & 78.957        \\ 
RPS(GP) & 0         &  0.250     & 38.206    & 37.715     & 2.056          & 78.228      \\ 
APS(TF) & 0         &  0.128     & 39.393    & 39.396     & 0.136          & 79.053       \\ 
RPS(TF) & 0         &  0.216     & 38.971    & 38.793     & 0.233          & 78.213      \\
\hline   
APS(GP) & 8         & 0.086     &  9.776     & 19.749     & 0.540          &  30.152    \\ 
RPS(GP) & 8         & 0.146     &  9.858     & 18.273     & 1.665          & 29.942    \\ 
APS(TF) & 8         & 0.036     & 10.026     & 20.061     & 0.037          & 30.160   \\ 
RPS(TF) & 8         & 0.104     & 9.992      & 19.700     & 0.119          & 29.914    \\
\hline   
APS(GP) & 12        & 0.162    & 7.715      & 14.740       & 1.007          & 23.625    \\ 
RPS(GP) & 12        & 0.240    & 7.731      & 13.703       & 1.926          & 23.599     \\ 
APS(TF) & 12        & 0.026    & 7.936      & 15.879       & 0.028          & 23.869  \\ 
RPS(TF) & 12        & 0.115    & 7.958      & 15.641       & 0.142          & 23.857    \\
\end{tabular}
\caption{
Contributions to the total energy for mixtures of BECs trapped in a disc geometry of $r_0=0$ and Corbino geometries with an inner radius of $r_0=8$ and $r_0=12$. For the disc trap $U_{1}=6000$, $U_2=5000$ and $U_{12}=6000$, for the broad Corbino with inner radius $r_0=8$, $U_{1}=7000$, $U_2=5000$ and  $U_{12}=6250$, and for the narrow Corbino with inner radius $r_0=12$, $U_{1}=7750$, $U_2=5000$ and  $U_{12}=6750$.
The radial phase separation is favoured for this imbalanced mixtures while the interface energy $E_s=E_{kin}+E^{12}_{int}$ for RPS is greater than its APS counterpart.}
\label{table:imbalanced}
\end{center}  
\end{table}
\par
We extend the results shown in Fig.~\ref{fig:ETFvsgamma0} by constructing a phase diagram as a function of the intercomponent interaction and the ratio of intracomponent interactions for all three trapping potentials considered. The boundaries between APS and RPS configurations are obtained by the TF approximation, and checked via the GP results.  As we discussed in Sec.~\ref{sec:balanced-case}, this agreement is due to the fact that the interface energy added to the TF approximation is necessary and sufficient to accurately predict the phase boundaries in the trap. To construct the APS-RPS phase boundary we take small increments for either the inter-component interaction $U_{12}$ or the intra-component interaction $U_1$ and monitor the APS and RPS state TF energies including the interface interaction. Whenever the ground state configuration changes we change direction from horizontal to vertical steps or vice versa.
When crossing the boundary we also check the GP energies and confirm that they are in agreement with the boundary crossing. Our step size for this construction is $\Delta (U_{12}/U_2)= \Delta (U_1/U_2) = 0.05$ and the TF and GP results are in agreement on this scale. At the end, we fit a line between the boundary crossing points to determine the phase boundary. The boundaries for the miscible state are calculated analytically according to the condition $U_1U_2 \leq U_{12}^2$. It is shown that this condition ensures miscibility against large deviations from uniformity~\cite{Ao1998, Pethick2008}.
\par
The phase diagrams reveal an interplay between the boundary effects and the imbalance in the intracomponent interaction in determining the the configuration of the phase separation. For a disc geometry in Fig.~\ref{fig:phasediagram}(a) the RPS is favourable in a large area of the phase diagram. Larger values of the intercomponent interactions move the boundary of the narrow APS region toward larger values of intracomponent interactions ratio. Larger intercomponent energy means larger difference between interface energies of  APS and RPS (see Fig.~\ref{fig:energies-symmetric}). Accordingly, the region for APS becomes larger moving from disc to Corbino with $r_0=8$ and more so for the narrower Corbino with $r_0=12$. On the other hand, the shape of the external potential, particularly its width in the case of the Corbino trap, defines the extent of the boundaries between two separated condensates. Eventually, an interplay between the boundary effects and the imbalance in the intracomponent interactions determines the configurations of the phase separations.
\section{Rotational properties}\label{sec:rotational-properites}
The velocity field of an atomic superfluid is defined by the gradient of the condensate phase~\cite{Pethick2008}, so that it is irrotational. This restriction on the velocity field of superfluids $\mathbf{v}$, i.e. $\nabla \times \mathbf{v}=0$, enforces the circulation around a closed path to be quantized. Accordingly, the velocity field of a vortex exhibits a profile with $1/r$ dependency perpendicular to the direction of applied rotational frequency (the experiments reveal only a $1/r$ dependency in the azimuthal component if the rotation is in the z-direction). This behaviour is completely different from a rigid body rotation which implies a linear dependency on $r$. Interestingly, for the APS scenario the quantization of the circulation breaks down~\cite{Shimodaira2010, White2016} and the azimuthal component of the velocity field with linear dependency on $r$ appears at the boundary regions~\cite{White2016} (for the RPS the circulation is quantized, similar to the miscible mixture).
Here, we investigate the shape of the velocity field of APS configurations for both Corbino and disk geometries.
\par
\begin{figure}[t]
\centering
\resizebox{.7\columnwidth}{!}{%
\includegraphics{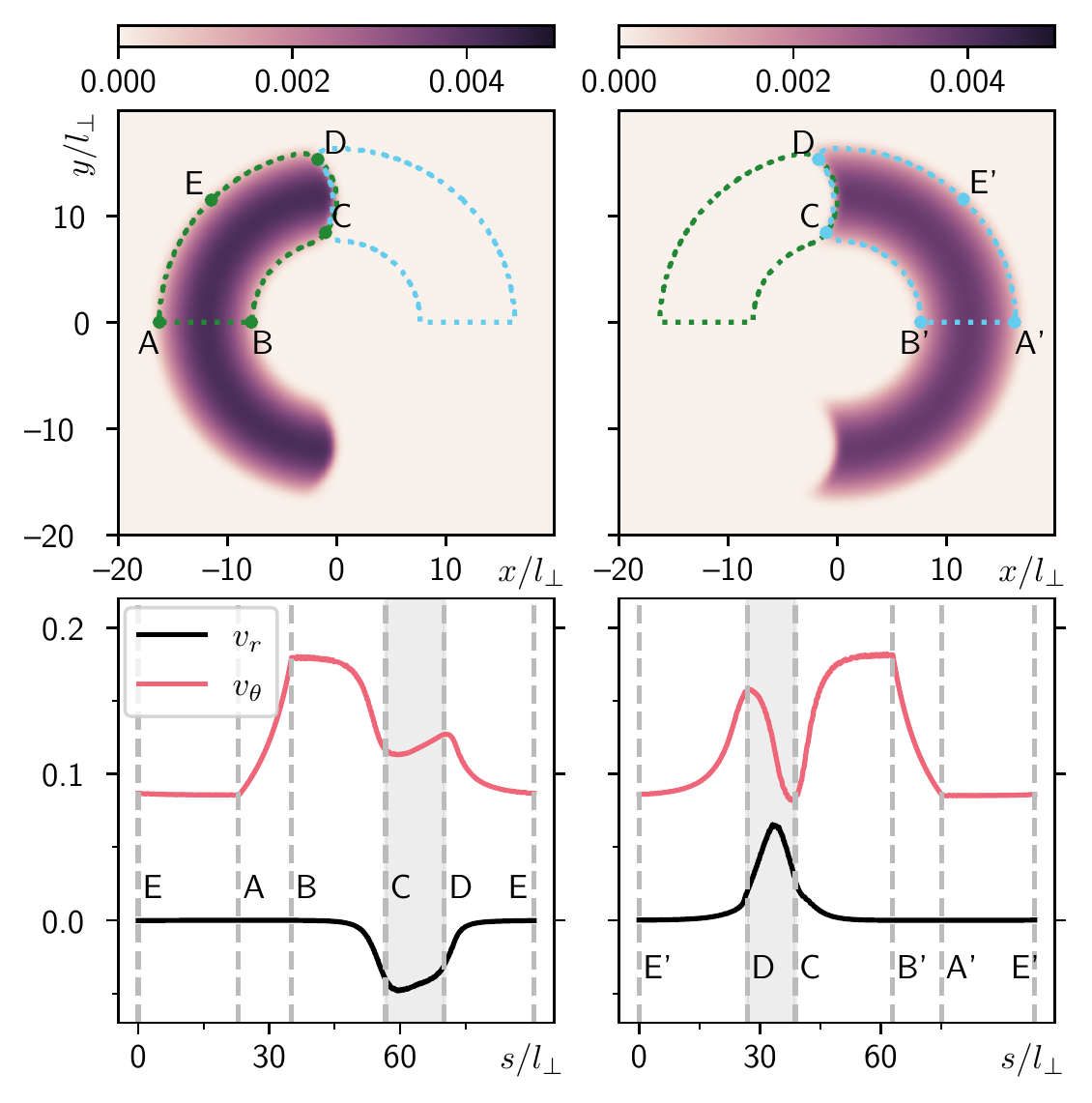}
}
\caption{Density (upper panel) and velocity component (lower panel) profiles of the APS configuration for an interaction-imbalanced mixture in a Corbino trap with $r_0=12$, $\Omega=0.01$, intracomponent interactions $U_2=3000$, $U_1=1.2U_2$, and intercomponent interaction $U_{12}=1.4U_2$. For both condensates (left and right panels), the azimuthal component $v_\theta$ of the velocity shows a superfluid flow in the bulk and a rigid body flow at the phase boundary (shaded area). 
\label{fig:velocity-corbino}
}
\end{figure}
\begin{figure}[t]
\centering
\resizebox{.7\columnwidth}{!}{%
\includegraphics{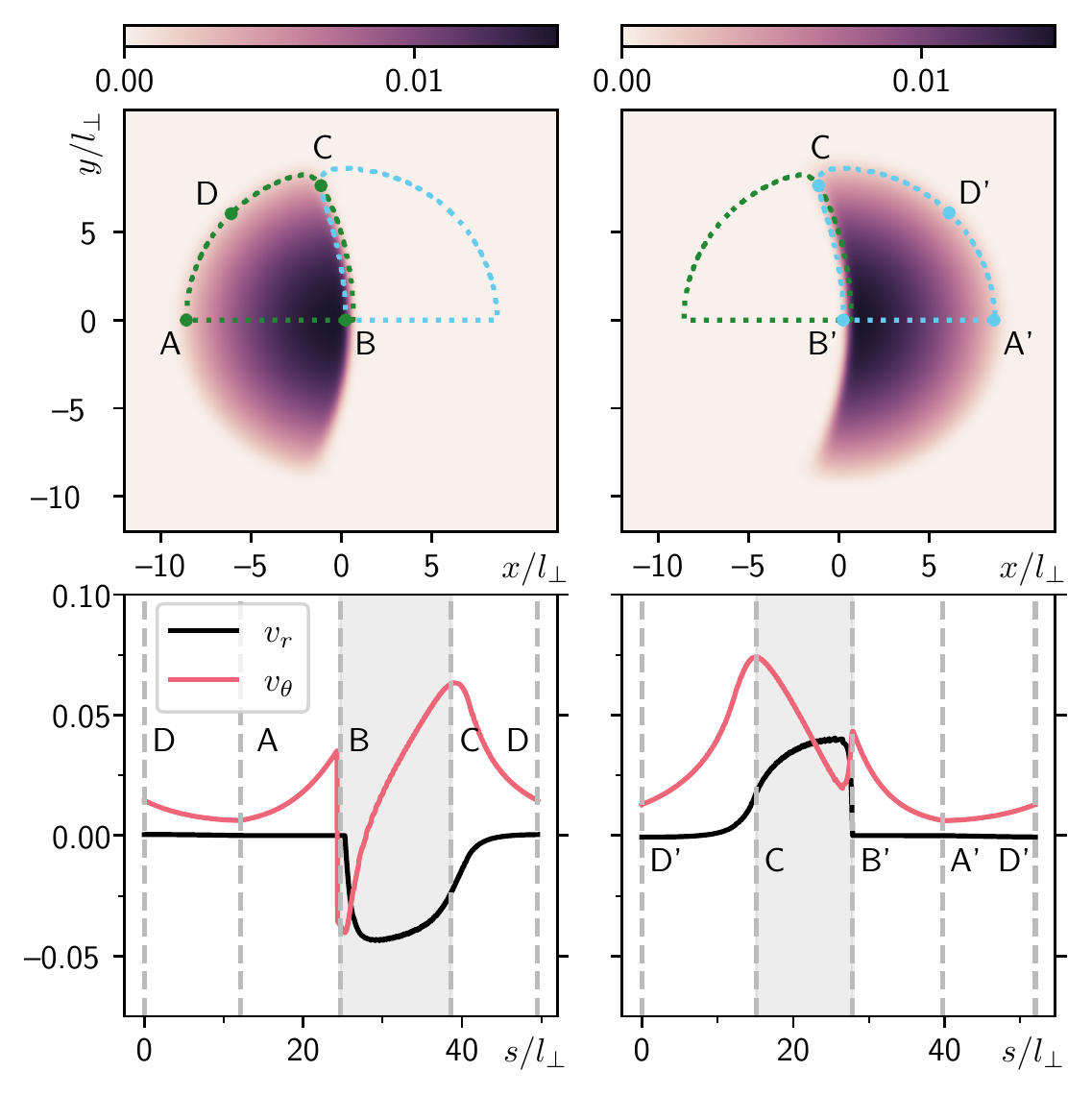}
}
\caption{Density (upper panel) and velocity component (lower panel) profiles of the APS configuration for an interaction-imbalanced mixture in a disc trap ($r_0=0$), with $\Omega=0.01$, intracomponent interactions $U_2=3000$, $U_1=1.025 U_2$, and intercomponent interaction $U_{12}=1.2U_2$. For both condensates (left and right panels), the azimuthal component of the velocity $v_\theta$ shows a superfluid flow in the bulk and a rigid body flow at the phase boundary (shaded area).
\label{fig:velocity-disc}
}
\end{figure}

We present the density distributions of immiscible mixtures in the APS configuration for Corbino ($r_0=12$) and disc geometries in upper panels of Fig.~\ref{fig:velocity-corbino} and Fig.~\ref{fig:velocity-disc}, respectively. In the lower panels we give the corresponding velocity field profiles. We follow a closed path of fixed density on each component's distribution starting from points E (E') in Corbino and D (D') in disc geometry. 
\par

The results for Corbino and disc geometries are qualitatively similar. The asymmetric interaction causes boundaries to curve in the density distributions, which are straight for interaction-balanced condensates (see top panel of Fig.~\ref{fig:density-distributions}). As seen from the velocity field profiles for both geometries (bottom panels of Fig.~\ref{fig:velocity-corbino} and Fig.~\ref{fig:velocity-disc}), the condensates display typical superfluid behaviour everywhere expect at the phase boundaries. The azimuthal component of the velocity, $v_\theta$, indicated by the solid red lines exhibits $1/r$ dependency as seen between points A (A') and B (B') for both condensates. However, $v_\theta$ becomes linear in $r$ between points C and D for the Corbino and B and C for the disc, where an extra non-zero radial velocity indicated by the solid red lines appears. The steep slope at points B and B' is the result of transition from superfluid flow in the bulk of condensate to the to classical rigid body flow along the interface boundary.
For the disc geometry the change happens near the origin over a smaller distance. Therefore, compared to the Corbino geometry where the change happens at a larger radial distance, the change in the flow profile is more pronounced for the disc geometry.

The radial velocities $v_r$ corresponding to each component in the mixture appear in opposite directions with similar magnitudes. Both profiles have been calculated with an imbalance in the intracomponent energies and they remain similar for the interaction-balanced case as shown in~\cite{White2016}. Unlike the RPS, in the APS the counter superfluid flow at the phase boundary can lead to instabilities and different excitations~\cite{Suzuki2010, Ishino2011, Ticknor2013, Eto2015}. Furthermore, for unequally charged systems a nonzero relative velocity at the phase boundary can lead to instabilities for both RPS and APS configurations.
\par
As can be inferred from the plots, the average angular velocity for equally charged condensates is non-zero 
indicating that the condensates have non-zero average angular velocity about the central axis. In contrast, we find that in a charged-uncharged mixture the average angular velocity of each component becomes zero and the uncharged component has constant phase in the mean-field approximation.
\par
\begin{figure}[t]
\centering
\resizebox{.5\columnwidth}{!}{%
\includegraphics{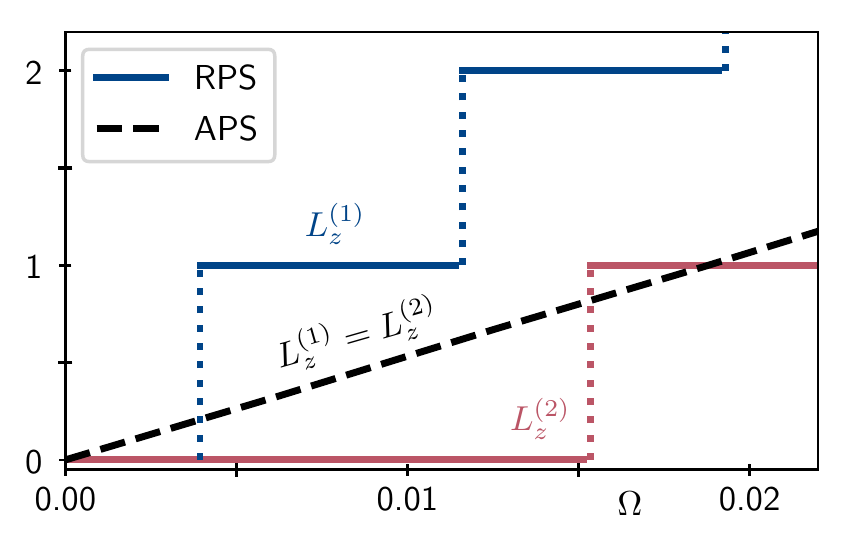}
}
\caption{Angular momenta $L_z^{(j)}$ of the condensates as a function of the applied rotational frequency $\Omega_1=\Omega_2=\Omega$ for Corbino ($r_0=8$) geometry. For the RPS configuration, the angular momenta are quantized. Here $U_1=7000$, $U_2=5000$, and $U_{12}=7000$. For the APS configuration the angular momenta become linear in $\Omega$ as the circulation quantization condition does not apply in this configuration. For this case $U_1=U_2=5000$, and $U_{12}=7000$.
\label{fig:angular-momentum}
}
\end{figure}

 
The breakdown of the circulation quantization in the APS manifests itself in the angular momentum properties of the system, which is seen in Fig.~\ref{fig:angular-momentum}. The angular momenta are calculated for a Corbino with $r_0=8$ and imbalanced intracomponent interactions as a function of rotation frequency/magnetic field. While the angular momentum of the RPS is quantized it becomes continuous for the APS. We limit our discussion to the slowly rotating limit such that the angular momentum per particle approximately equal to less than $\hbar$.

\par
In the mean field GP and TF approaches, where only density-density interactions are considered, the phases of the superfluid wave functions do not enter the interactions. The absence of these phases prevents any sort of interplay between the velocity fields of the superfluids. Therefore, at the GP and TF mean-field level it is not possible to observe an angular momentum transfer between two unequally charged superfluids~\cite{Aksu2019}.
\section{Summary and conclusion}\label{sec:Conclusion}
We studied the phase separation configurations and their rotational properties for a mixture of two interacting charged Bose-Einstein condensates subject to a magnetic field trapped in two different geometries; disc and Corbino. The azimuthal phase separation and the radial phase separation are two types of phase separation configurations that occur for such mixtures. 
\par 
In order to determine the phase separation configurations we calculated the ground state energies of the configurations using the Gross-Pitaevskii and the Thomas-Fermi approximations. We modified the Thomas-Fermi approximation for the phase separated scenarios, and added the contribution of the boundary effects to our Thomas-Fermi approach. We showed that the results of the modified Thomas-Fermi approximation are in good agreement with those obtained from the Gross-Pitaevksii approach, and can be used in determining the configurations of the phase separations. We obtained a phase diagram exhibiting the range of these configurations as a function of the inter and intracomponent interactions. 
\par
The phase separation configurations are determined by the imbalance of the intracomponent interactions and the shape of external potential~\cite{Svidzinsky2003, Shimodaira2010, Mason2013, Abad2014}. We show that the APS is the only ground state of a mixture with all equal physical parameters, even for very large intracomponent interactions. The geometry of the trap does not play a role in such a symmetric mixture of BECs. In this case, with both components enjoying the same density distributions and intracomponent interaction strengths, the mixture tends to minimize the contribution of the boundary effects to the total energy. 
\par
We showed that a phase transition from APS to RPS occurs by introducing an imbalance in the system~\cite{Abad2014}. This transition occurs in order to minimize the total energy through which the condensate with a larger intracomponent energy moves radially outwards with lower density satisfying the pressure balance at the phase boundary. The configuration of the phase separation in this case is determined by an interplay between the interface energy and the intracomponent imbalance. While the radial phase separation is widely favoured in disc geometry, the azimuthal phase separation is favoured for narrower Corbino geometries. 
\par
We explored the rotational properties of the spatially separated condensates under the magnetic field studying their angular momenta and velocity fields. We showed that the circulation condition breaks down during the azimuthal phase separation. For charged-imbalanced mixtures, the rotational properties show a qualitative difference even in the mean-field level in that the rotation about the central axis stops for a charged-uncharged mixture. The transfer of angular momentum between the components provides an interesting area of further research. For unequally charged systems the difference in relative velocities can lead to instabilities for both RPS and APS configurations. Beyond-mean-field treatments are needed for this purpose. The mean-field solutions in this work provide the starting point for such investigations.
\par
In typical BEC mixture experiments different hyperfine states of $^{87}$Rb \cite{Hall1998, Mertes2007} are considered. In these experiments for a pancake shaped two-dimensional geometry, the radial and axial trapping frequencies can be $\omega_\perp = 2\pi \times 30\,$Hz and $\omega_z = 2\pi \times 85\,$Hz~\cite{Mertes2007}, respectively, satisfying the condition $\omega_\perp \ll \omega_z$. Taking into account equal populations of particles in the range $N= 4.5 \times 10^4 - 6.5 \times 10^5$, we get inner-component interactions in the range $U_1 = U_2 = U = [1000 - 15000]$, using approximate s-wave scattering length $a_s \approx 5\,$nm. The system can be experimentally driven from a miscible phase to an immiscible phase by adjusting the Feshbach resonance, and thus the possibility of working in different phases.

\ack{
This work is supported by T\"{U}B\.{I}TAK under Project No. 117F469. A.L.S. acknowledges the hospitality of the Center for Non-linear Studies (CNLS).
}

\section*{Refenrences}

\bibliographystyle{iopart-num}
\bibliography{BEC}

\end{document}